\documentclass[aps,prb,twocolumn,showpacs,superscriptaddress]{revtex4}

\usepackage{graphicx}
\usepackage{amsmath}
\usepackage{amssymb}
\usepackage{lscape}
\usepackage{braket}
\usepackage{float,hyperref,epstopdf}
\DeclareMathOperator{\Tr}{Tr}

\begin{document}
\title{Order-from-disorder and critical scalings of spontaneous magnetization \\
 in random-field quantum spin systems}

\author{Anindita Bera}
\affiliation{Department of Applied Mathematics, University of Calcutta, 92, A.P.C. Road, Kolkata 700 009, India}
\affiliation{Harish-Chandra Research Institute, Chhatnag Road, Jhunsi, Allahabad 211 019, India}
\author{Debraj Rakshit}
\affiliation{Harish-Chandra Research Institute, Chhatnag Road, Jhunsi, Allahabad 211 019, India}
\author{Maciej Lewenstein}
\affiliation{ICFO-Institut de Ci$\grave{e}$ncies Fot$\grave{o}$niques, Av. C.F. Gauss 3, 08860 Castelldefels (Barcelona), Spain}
\affiliation{ICREA-Instituci{\'o} Catalana de Recerca i Estudis Avancats, Lluis Companys 23, 08010 Barcelona, Spain}
\author{Aditi Sen(De)}
\affiliation{Harish-Chandra Research Institute, Chhatnag Road, Jhunsi, Allahabad 211 019, India}
\author{Ujjwal Sen}
\affiliation{Harish-Chandra Research Institute, Chhatnag Road, Jhunsi, Allahabad 211 019, India}
\author{Jan Wehr}
\affiliation{Department of Mathematics, University of Arizona, Tucson, AZ 85721-0089, USA}

\date{\today}
\begin{abstract}
We investigate the effect of a unidirectional quenched random field on the anisotropic quantum spin-1/2 $XY$ model, 
which magnetizes spontaneously in the absence of the random field. We adopt mean-field approach to show that spontaneous magnetization persists even in the presence of 
this random field but the magnitude of magnetization gets suppressed due to disorder, and the system magnetizes in the directions parallel and transverse to the random field. 
Our results are obtained by analytical calculations within perturbative framework and by numerical simulations. Interestingly, we show that it is possible to enhance 
a component of the magnetization in presence of the disorder field provided we apply an additional constant field in the $XY$ plane. Moreover, we derive generalized expressions for the critical 
temperature and the scalings of the magnetization near the critical point for the $XY$ spin system with arbitrary fixed quantum spin angular momentum. 

\end{abstract}

\maketitle

\section{Introduction}
\label{sec_introduction}
Disorder is ubiquitous in solid state materials. The effectd of disorder are often non-intuitive, and hence have generated a lot of attention in condensed-matter physics \cite{dis2, dis3, anderson,ahufinger}. 
Disorder can currently also be engineered in a controlled way via ultracold atoms trapped in optical lattices subjected to an additional, e.g. optical speckle potential \cite{ahufinger}. 
Disordered systems are often endowed with nontrivial properties, dramatically different from those of their homogeneous counterparts. The novel quantum phases \cite{Mezard,yao} and  unique phenomena, 
such as Anderson localization \cite{anderson2}, dynamical many-body localization  \cite{mbl} and presence/absence of thermalization \cite{Naturephys-gogolin}, and high $T_c$ superconductivity \cite{Auerbach} are some 
of the prominent examples. Quenched disorder in type II superconductors have been investigated by using Ginzburg-Landau theory
to uncover ``glassy" properties in the system \cite{GL}.
In particular, considerable efforts have been dedicated to understanding the effects of disorder in spin models, both classical and quantum \cite{Nagaoka,wehr-1,wehr-2,classical,group}.  

In classical systems with continuous symmetry, it has been shown that a random field with the symmetry of the system may cause significant changes in its properties \cite{imry,imbrie}.  A small random magnetic field of this kind can destroy magnetization in a classical spin system at any temperature, including zero-temperature.  E.g., two-dimensional $XY$ and Heisenberg models do not magnetize in presence of random fields with $SO(2)$ and $SO(3)$ symmetries respectively \cite{imry, wehr-1} (see also \cite{imbrie} for the analogous effect in systems with discrete symmetry). The effect prevails in higher dimensions as long as the random field exhibits the corresponding symmetry \cite{wehr-1}. However, in absence of appropriate symmetry, these systems exhibit spontaneous magnetization \cite{wehr-2,classical}. 

Despite the difficulty in dealing with spontaneous magnetization and other
system characteristics in quantum disordered systems, many 
important results have been obtained. A quantum Hall nematic phase has been predicted
in a zero temperature two-dimensional electron system that is unstable to
weak disorder \cite{ek, paanch}. Collective properties of magnetic
impurities on a topological surface were studied both theoretically
\cite{tin} and experimentally \cite{char}. Experimental and theoretical
investigations in solid state systems have revealed that alloy disorder can
reduce the Curie temperature in the system \cite{dui}. Arbitrarily weak
interparticle interactions were shown to destabilize the surface states of
topological superconductors in the presence of non-magnetic disorder
\cite{chhoi}. The critical behavior and effective exponents in
ferromagnetic quantum phase transitions of disordered systems were derived
in Ref.\cite{sath}.

Recently, 
a mean-field classical spin model with $SO(n)$ symmetry was considered and its
spontaneous magnetization was investigated 
in presence of unidirectional random fields \cite{classical}.
The natural question is how a symmetry breaking random field  affects these systems in the quantum limit. As already mentioned before, this question is particularly relevant due to the current accessibility of disordered quantum spin models in experiments \cite{ahufinger}. If we restrict ourselves to the $XY$ model with a random field, and to one dimension, then such quantum system of a moderately large size can be investigated using the Jordan-Wigner technique \cite{jordon}. However, higher dimensional quantum spin models remain intractable due to the lack of analytical and numerical techniques, even for ordered systems. We work within the mean-field (MF) approximation which is often effective in capturing system's properties qualitatively. Numerical schemes like exact diagonalization are usually inefficient for even moderately large systems, due to the exponential growth of the dimension of the system's Hilbert space.   Metastability effects and slow relaxation rates, usually present in disordered systems, make other numerical simulation techniques such as density-matrix renormalization group, Monte-Carlo approach difficult to apply, particularly in higher-dimensional lattice systems with higher-dimensional spins.  The mean-field approach essentially liberates one from these immediate challenges allowing for a detailed analysis, that enables us to answer some of the key questions. Of course, the price to pay is that the mean field approach is not expected to describe the details of the critical behavior precisely, except in high dimensions.

In this work, we consider the quantum spin-$1/2$ $XY$ model with anisotropic interaction in presence of a unidirectional quenched random field. The purely isotropic case, i.e., 
the quantum $XY$ model with vanishing anisotropy parameter and vanishing disorder, exhibits a spontaneous magnetization which has circular symmetry. The anisotropy breaks the continuous symmetry even for the pure system. The pure spin-$1/2$ $XY$ system magnetizes below a certain critical temperature.  This system still magnetizes when a random field is introduced at a critical temperature which is higher than in the system without disorder. 
We show by means of numerical as well as perturbative analysis that the system now magnetizes in specific directions, which is either along the parallel or the perpendicular directions to the random field. The critical temperature in both cases decreases with the increase of the random field strength.
We find that the critical temperature to magnetize in transverse and parallel directions show opposite behavior with respect to the anisotropy in the system. 
Specifically, with the increase of anisotropy, the critical temperature corresponding to the transverse magnetization increases, while the opposite happens for parallel magetization.
It is important to mention that for vanishing anisotropy parameter, the continuous symmetry of the pure $XY$ system is broken by the 
introduction of an arbitrarily small random field. We also present general expressions of the scalings of critical temperature of magnetization for the  quantum $XY$ spin systems with arbitrary half-integer and integer spins. 
 
In addition, adding a constant magnetic field along with the random field, we find that the component of the magnetization perpendicular to the random field gets enhanced due to the disorder, an effect known as ``random-field-induced order" or ``order from disorder", which has also been reported for several other models \cite{odd-dis,group}. 

The rest of the paper is arranged as follows. In Sec.~\ref{model}, we introduce the spin-$1/2$ quantum $XY$ model in presence of the random field, and subsequently derive critical scaling of the magnetization via perturbative approach. We also discuss numerical results obtained within the MF approximation. In Sec.~\ref{steady}, we demonstrate, both numerically and analytically, the order from disorder phenomena under the influence of an additional constant field. In Sec.~\ref{general}, we derive the generalized expressions for the case of arbitrary integer and half-integer spins. Finally, we conclude in Sec.~\ref{summary}.
                                                                                                                                                                                                                                                                                                                                                                                                                                                                                                                                                                                                                                                                                                                                                                                                                                                                                                                                                                                                                        
\section{QUANTUM SPIN-1/2 XY MODEL IN A RANDOM FIELD}
\label{model}
We consider the quantum spin-$1/2$  $XY$ model in a random field. Our aim is to study the effect of the random field on the magnetization as a function of temperature, and to find the scaling of the magnetization around the critical temperature. In the following subsection, we introduce the system and the mean-field approximation.
\subsection{The system and its mean-field treatment}
The Hamiltonian of the ferromagnetic quantum $XY$ model is given by
\begin{equation}
\label{ham}
H_{XY} = H_{int}+H_{ext},
\end{equation}
where
\begin{equation}
\label{hint}
H_{int}=-\sum_{|i-j|=1}^N[J_x \sigma_i^x \sigma_{j}^x+J_y \sigma_i^y \sigma_{j}^y].
\end{equation}
The coupling constants, $J_{\alpha}$, are assumed to be positive. They can be further expressed in terms of an anisotropy parameter, $\gamma$, as $J_x=J (1+\gamma)$ and $J_y=J (1-\gamma)$. The indices, $i$ and $j$ denote the sites of an arbitrary $d$ dimensional lattice and  $\sigma_i^\alpha, \alpha=x, y$ are the Pauli matrices on the $i^{th}$ site. The part of the Hamiltonian in Eq.~(\ref{ham}) due to an inhomogeneous magnetic field, $H_{ext}$, equals
\begin{equation}
\label{hext}
H_{ext}=-\epsilon \sum_{i} \vec{h}_i\cdot\vec{\sigma}_i,
\end{equation}
 where $\epsilon$ $(>0)$, a dimensionless parameter that quantifies the strength of the randomness, is typically chosen to be small. The unidirectional random field, $\vec{h}_i$,  is chosen to be $\vec{h}_i=\eta_i\cdot\hat{e}_y$, where $\eta_{i}$ are independent and identically distributed quenched Gaussian random variables with zero mean and unit variance, and $\hat{e}_y$ is the unit vector in the $y$-direction. Within the mean-field limit, as we show below, the pure systems governed by the Hamiltonian $H_{int}$ magnetizes even at low-dimensions.  This does not contradict the Mermin-Wagner-Hohenberg theorem \cite{Mermin}, which predicts no spontaneous magnetization in one- and two-dimensions since the predictions made by the  mean-field approximation only becomes accurate in higher dimensions \cite{Justin} where Mermin-Wagner-Hohenberg theorem is not valid. Interestingly, it has been shown that a uniaxial random field may help the system to magnetize even at two-dimension \cite{wehr-2,Crawford}. Note that had the random field, $\vec{h}_i$, been chosen to be invariant under rotations, the system would not magnetize at any nonzero temperature in any dimension $d \le 4$ \cite{wehr-1,imry,imbrie}.  \\

Within the mean-field approximation, each spin is regarded to be reacting to an average field due to all the other spins in the system. Assuming $N$ to be the total number of spins in the system, the effective interaction, replacing the nearest neighbor interaction in $H_{int}$, for large $N$, equals approximately $H_{int}=-1/N(\sum_{\alpha=x,y}\sum_{j;j\ne i}^N \nu J_{\alpha}' \sigma_i^{\alpha} )\sigma_{j}^{\alpha}=-\sum_{\alpha=x,y} \nu J_{\alpha}' m_{\alpha} \sigma_i^{\alpha} $, where $m_{\alpha}=\frac{1}{N} \sum_{j=1}^N \sigma_{j}^{\alpha}$. $J'$ is the coupling constant and $\nu$ is the coordination number which depends on the geometry of the lattice.
Note that mean-field approximation provides close to the exact description for large dimensional lattice systems \cite{Thompson}.  Within the mean-field approximation, the Hamiltonian, $H_{XY}$, is given by
\begin{equation}
\label{xy_disorder1}
H=-J (1+\gamma) m_x \sigma_x-J (1-\gamma) m_y \sigma_y-\epsilon \vec{\eta}\cdot\vec{\sigma},
\end{equation}
where the operators $m_\alpha$ are replaced by their average values denoted by the same symbol, in the canonical equilibrium state at absolute temperature $T$ and $J=J' \nu$. Now in order to monitor the behavior of the  magnetization as a function of temperature, one needs to calculate the expectation value of the spin operators, $\sigma_{\alpha}$, $\alpha=x,y$. So in the mean-field regime, the magnetization of the system governed by the Hamiltonian, $H$ (see Eq.~(\ref{xy_disorder1})), is given by 
\begin{equation}
\label{xy_disorder}
m_{\alpha}={\rm Av}_\eta\left[\frac{Tr\left[\sigma_{\alpha} \exp(-\beta H)\right]}{Tr\left[\exp(-\beta H)\right]}\right],
\end{equation}
where $\beta=1/(k_B T)$, where $k_B $ is the Boltzmann constant, and ${\rm Av}_\eta[.]$ denotes the  average over the realizations of randomness. From Eqs.~(\ref{xy_disorder1}) and (\ref{xy_disorder}), we obtain a coupled set of the following two equations:
\begin{figure}[t]
\vspace*{+.2cm}
\includegraphics[angle=0, width=65mm]{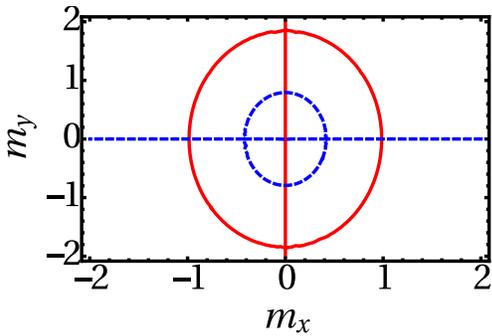}
\caption{(Color online) System magnetizes in directions parallel and transverse to the disorder field. Zero contour lines of the $F_x^{\epsilon,2}(m)$ and $F_y^{\epsilon,2}(m)$ in Eqs. (\ref{xy_disorder6}) [solid-red] and (\ref{xy_disorder7}) [dotted-blue] for $\gamma=0.3$, $\epsilon/J=0.1$ and  $J \beta=2$, respectively, as functions of $m_x$ and $m_y$. All quantities are dimensionless.}
\label{fig_contour}
\end{figure}
\begin{equation}
\label{xy_disorder2}
m_\perp^{\epsilon,2}\equiv m_x={\rm Av}_\eta\left[\frac{J (1+\gamma) m_x}{k^\epsilon} \tanh(\beta k^\epsilon)\right]
\end{equation} 
and
\begin{equation}
\label{xy_disorder3}
m_\parallel^{\epsilon,2}\equiv m_y={\rm Av}_\eta\left[\frac{J (1-\gamma) m_y+\epsilon \eta}{k^\epsilon} \tanh(\beta k^\epsilon)\right],
\end{equation}
where $k^\epsilon=\sqrt{(J (1+\gamma) m_x)^2+(\epsilon \eta+J (1-\gamma) m_y)^2}$. Note that  the subscripts $\perp$ and $\parallel$ classify two distinct cases, as would be apparent later. The superscripts $\epsilon$ and $2$ keep track of the strength of the disorder and of the value of the spin ($1/2$) respectively. 
\subsection{Critical point and scaling of magnetization near criticality}
\label{near}
The magnetization, $\vec{m}$, can be obtained by finding the common zeros of the following two functions obtained from Eqs.~(\ref{xy_disorder2}) and (\ref{xy_disorder3}))
\begin{equation}
\label{xy_disorder4}
F_x^{\epsilon,2}(\vec{m})={\rm Av}_\eta[\frac{J (1+\gamma) m_x}{k^\epsilon} \tanh(\beta k^\epsilon)]-m_x,
\end{equation}
and
\begin{equation}
\label{xy_disorder5}
F_y^{\epsilon,2}(\vec{m})={\rm Av}_\eta[\frac{J (1-\gamma) m_y+\epsilon \eta}{k^\epsilon} \tanh(\beta k^\epsilon)]-m_y,
\end{equation}
where $k^\epsilon=\sqrt{(J (1+\gamma) m_x)^2+(\epsilon \eta+J (1-\gamma) m_y)^2}$. Let us set $m_x=m \cos\phi_1$, $m_y=m \sin\phi_1$, and $\vec{m}=(m_x,m_y)$.
\\

By performing perturbative analysis, we can study the magnetization for small $\epsilon$. Taylor series expansion of Eqs. (\ref{xy_disorder4}) and (\ref{xy_disorder5})  in $\epsilon$ around $\epsilon=0$ gives
\begin{equation}
\label{xy_disorder6}
F_x^{\epsilon,2}(\vec{m})=c_1+\frac{1}{2} b_1 \epsilon^2+O(\epsilon^4),
\end{equation}
and
\begin{equation}
\label{xy_disorder7}
F_y^{\epsilon,2}(\vec{m})=c_2+\frac{1}{2} b_2 \epsilon^2+O(\epsilon^4),
\end{equation}
where
\begin{eqnarray}
\label{xy_disorder8}
c_1= m_x (-1+\frac{J (1+\gamma)}{k} \tanh(\beta k)),
\end{eqnarray}
\begin{eqnarray}
\label{xy_disorder10}
c_2=m_y (-1+\frac{J (1-\gamma)}{k} \tanh[\beta k]),
\end{eqnarray}
\begin{widetext}
\begin{eqnarray}
\label{xy_disorder9}
b_1= \frac{-3 J^3 m_x m_y^2 \beta (1-\gamma)^2 (1+\gamma)}{k^4} \frac{1}{\cosh[\beta k]^2}
+\frac{J m_x \beta (1+\gamma)}{k^2} \frac{1}{\cosh[\beta k]^2}
+\frac{3 J^3 m_x m_y^2 (1-\gamma)^2 (1+\gamma) \tanh[\beta k]}{k^5}
\nonumber\\
-\frac{J m_x (1+\gamma) \tanh[\beta k]}{k^3}
-\frac{(2 J^3 m_x m_y^2 \beta^2 (1-\gamma)^2 (1+\gamma) \tanh[\beta k])}{k^3} \frac{1}{\cosh[\beta k]^2},
\end{eqnarray}
\begin{eqnarray}
\label{xy_disorder11}
b_2=\frac{-3 J^3 m_y^3 \beta (1-\gamma)^3}{k^4} \frac{1}{\cosh[\beta k]^2}
+\frac{3 J m_y \beta (1-\gamma)}{k^2} \frac{1}{\cosh[\beta k]^2}
+\frac{3 J^3 m_y^3 (1-\gamma)^3 \tanh[\beta k]}{k^5}
\nonumber\\
-\frac{3 J m_y (1-\gamma) \tanh[\beta k]}{k^3}
-\frac{2 J^3 m_y^3 \beta^2 (1-\gamma)^3 \tanh[\beta k]}{k^3} \frac{1}{\cosh[\beta k]^2}.
\end{eqnarray}
\end{widetext}
Here $k=\sqrt{(J (1+\gamma) m_x)^2+(J (1-\gamma) m_y)^2}$.

A contour analysis at this point becomes helpful to characterize the behavior of the system, in particular, in finding the directions in which the system magnetizes (see Fig.~\ref{fig_contour}). This amounts to identification of the zero-contour lines corresponding to Eqs.~(\ref{xy_disorder6}) and (\ref{xy_disorder7}). The intersection points of the zero-contour lines are possible solutions of the magnetization. For any given set of parameters, one immediately finds that the roots of the $F_x^{\epsilon,2}(\vec{m})$ and $F_y^{\epsilon,2}(\vec{m})$ exist only at $\phi_1=0$ or $\pi/2$. This implies that the magetization is either transverse to the random field (case I) or parallel to the random field (case II). Note that for $\epsilon=0$ and $\gamma=0$, the zero contour lines for both the equations would lie on top of each other due to the circular symmetry in the system. However, an arbitrary small random field is enough to break this symmetry. It follows from the contour analysis that above a certain temperature, the critical temperature, the zero-contour lines corresponding to Eqs.~(\ref{xy_disorder6}) and (\ref{xy_disorder7}) intersect only if $m_x=m_y=0$, which is a trivial solution. 

In order to find the critical temperature and the scaling of magnetization near criticality, we perform another round of Taylor expansions in Eqs.~(\ref{xy_disorder6}) and (\ref{xy_disorder7}) around $m=0$ to obtain
\begin{eqnarray}
\label{xy_disorder12}
F_x^{\epsilon,2}(\vec{m})=-\frac{1}{3} (3+J \beta (1+\gamma) (-3+\beta^2 \epsilon^2)) m \cos\phi_1
\nonumber\\
-\frac{1}{3!} \frac{2}{5} J^3 \beta^3 (1+\gamma)^3 (5-8 \beta^2 \epsilon^2+4 \beta^2 \epsilon^2 \cos{2 \phi_1}) m^3 \cos\phi_1
\nonumber\\
+O(m^5),\nonumber\\
\end{eqnarray}
and
\begin{eqnarray}
\label{xy_disorder13}
F_y^{\epsilon,2}(\vec{m})=(-1+J \beta (1-\gamma) (1-\beta^2 \epsilon^2)) m \sin\phi_1
\nonumber\\
-\frac{1}{3!} \frac{2}{5} J^3 \beta^3 (1-\gamma)^3 (5-16 \beta^2 \epsilon^2+4 \beta^2 \epsilon^2 \cos{2 \phi_1}) m^3 \sin\phi_1
\nonumber\\
+O(m^5).
\nonumber\\
\end{eqnarray}
The contour analysis implies that the allowed values of $\phi_1$ are $\pi/2$ and $0$.  For transverse magnetization, i.e., for the case I with $\phi_1=0$,  $F_y^{\epsilon,2}(\vec{m})$ vanishes (see Eq.~(\ref{xy_disorder13})). The nontrivial solutions, which solely appear from Eq.~(\ref{xy_disorder12}), are given by 
\begin{eqnarray}
\label{xy_disorder14}
m_\perp^{\epsilon,2}=\pm \sqrt{5} \sqrt{\frac{3+J \beta (1+\gamma) (\epsilon^2 \beta^2-3)}{(-5+4 \epsilon^2 \beta^2) J^3 \beta^3 (1+\gamma)^3}}.
\end{eqnarray}
The critical point is obtained by setting $m_\perp^{\epsilon,2}=0$ in  Eq.~(\ref{xy_disorder14}). We get
\begin{equation}
\label{xy_disorder14-1}
\beta_{c,\perp}^{\epsilon,2}=\frac{1}{J (1+\gamma)}+\frac{\epsilon^2}{3 J^3 (1+\gamma)^3}.
\end{equation}
Here $\beta$ is associated with the subscript $\perp$, following the similar convention in magnetization.

The magnetization values corresponding to case II are obtained by setting $\phi_1=\pi/2$ in Eqs.~(\ref{xy_disorder12}) and (\ref{xy_disorder13}). In this case, the function in Eq.~(\ref{xy_disorder12}) vanishes. The nontrivial solutions of Eq.~(\ref{xy_disorder13}) are given by
\begin{equation}
\label{xy_disorder16}
m_\parallel^{\epsilon,2}=\pm \sqrt{3} \sqrt{\frac{1+J \beta (1-\gamma) (\epsilon^2 \beta^2-1)}{J^3 \beta^3 (1-\gamma)^3 (4 \epsilon^2 \beta^2-1)}}.
\end{equation}
Subjecting Eq.~(\ref{xy_disorder16}) to the constraint $m_\parallel^{\epsilon,2}=0$, we obtain the following expression for the critical temperature:
\begin{equation}
\label{xy_disorder16-1}
\beta_{c,\parallel}^{\epsilon,2}=\frac{1}{J (1-\gamma)}+\frac{\epsilon^2}{J^3 (1-\gamma)^3}.
\end{equation}
Following the set of Eqs.~(\ref{xy_disorder14})-(\ref{xy_disorder14-1})) and (\ref{xy_disorder16})-(\ref{xy_disorder16-1})), one can immediately infer that the effect of disorder is more conspicuous if the system chooses to magnetize along the direction parallel to the random field, as compared to the other possibiliy with a transverse magnetization. Interestingly, the findings are consistent with the pictures drawn within a classical limit \cite{classical} for the isotropic case i.e., $\gamma=0$. However, quantitative comparison shows that the analysis with classical spins overestimates the effect of disorder on the critical scaling.

Note that one can immediately deduce the scaling expressions for the magnetizations and critical temperature of the isotropic ordered systems by putting $\gamma=\epsilon=0$ in Eqs.~(\ref{xy_disorder16}) and (\ref{xy_disorder16-1}) (or equivalently, in Eqs.~(\ref{xy_disorder14}) and (\ref{xy_disorder14-1})). In this case, the solutions form a circle in the $XY$ plane. This can be easily understood by following the set of Eqs.~(\ref{xy_disorder2}) and (\ref{xy_disorder3}), which become identical for the isotropic ordered systems. The symmetry is broken in presence of the random field and then the system prefers a specific direction of magnetization. 

Moreover, it is clear from Eqs.~(\ref{xy_disorder14-1}) and (\ref{xy_disorder16-1}), that critical temperature at which system magnetizes increases with the coordination number, i.e., with the dimension of the system, for any value of $\gamma$.

\begin{figure}[t]
\includegraphics[angle=0, width=65mm]{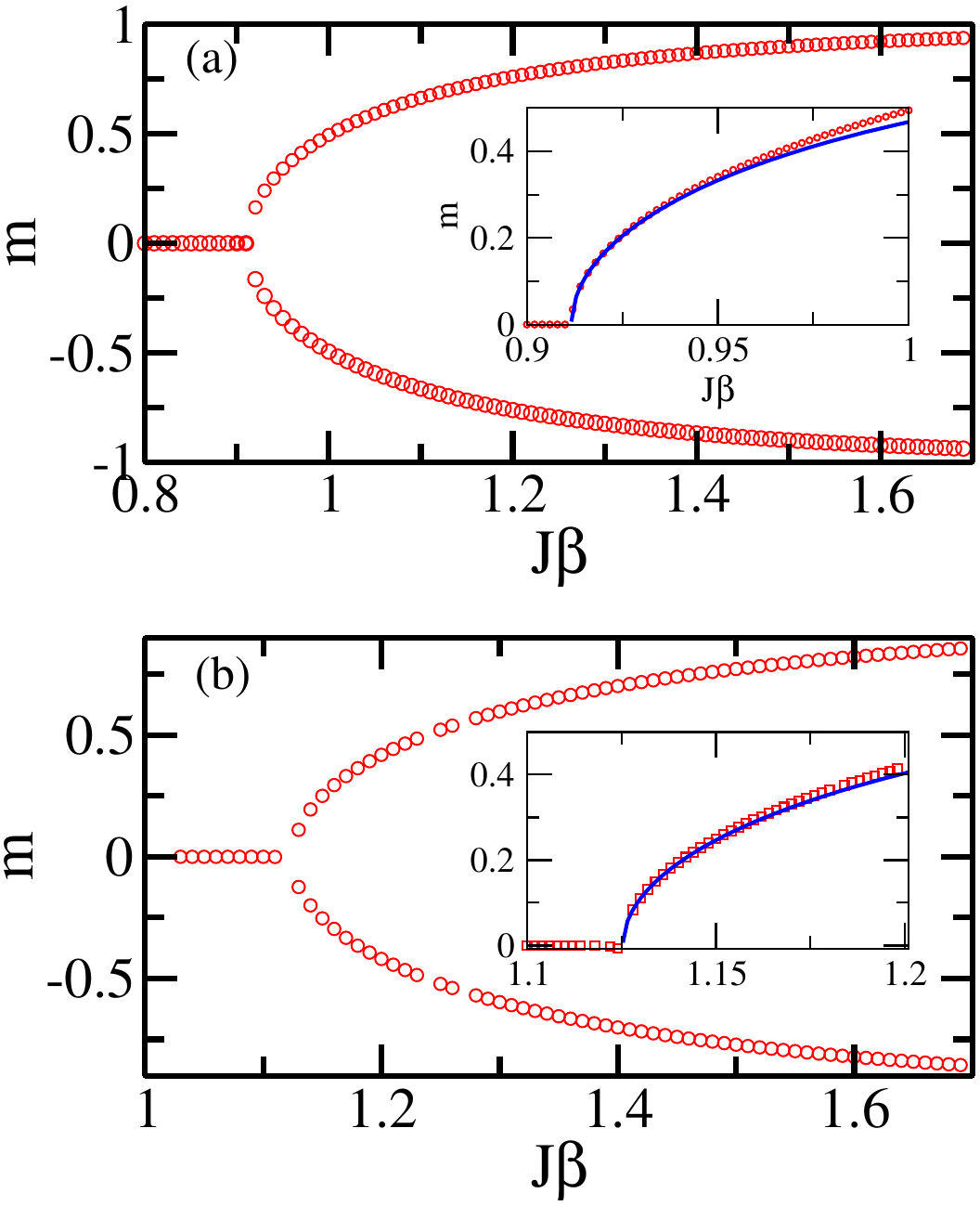}
\caption{(color online) Numerical and analytical results exhibit persistence of spontaneous magnetization in specific directions even after insertion of disorder. Numerical results for the magnetization as a function of $J \beta$, in the directions a) transverse and b) parallel to the disordered field. Red  circles correspond to the roots of Eqs.~(\ref{xy_disorder2}) and (\ref{xy_disorder3}) with $\epsilon/J=0.1$ and $\gamma=0.1$. Insets: The blue solid lines correspond to the analytic solutions derived for small $m$ given in Eqs.~(\ref{xy_disorder14}) and (\ref{xy_disorder16}) for the same set of parameters. The red circles are the numerical results. We find that the numerical and analytical results agree in the small $m$ regime. All quantities are dimensionless.}
\label{fig_epsilon}
\end{figure}
\begin{figure}[t]
\includegraphics[angle=0, width=65mm]{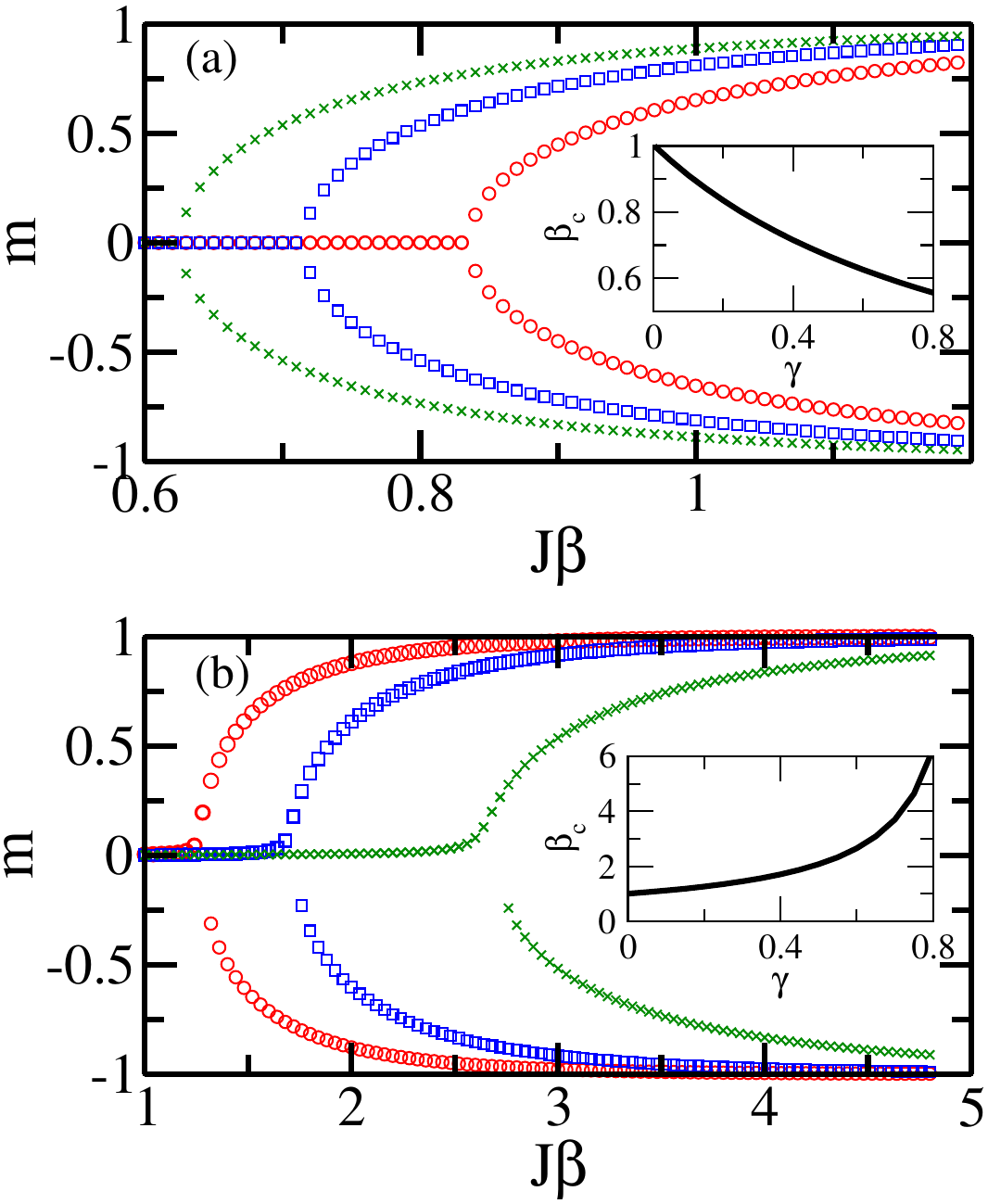}
\caption{(color online)  Magnetization as a function of $J \beta$ for different choices of the anisotropy constant, $\gamma$, in the directions a) transverse and b) parallel to the random field. Circles, squares, and crosses correspond to the magnetization of the system with $\epsilon/J=0.1$ and $\gamma=0.2$, 0.4, and 0.6, respectively. Insets show the inverse critical temperatures as functions of $\gamma$ for $\epsilon/J=0.1$. All quantities are dimensionless.}
\label{fig_epsilon_1}
\end{figure}
\subsection{Numerical results}
In the previous section, we derived the expressions for magnetization near the critical point. However, away from the critical point, where the perturbative approach is no longer valid, one has to rely on a numerical simulation to find the roots of the coupled set of equations, given in Eqs.~(\ref{xy_disorder2}) and (\ref{xy_disorder3}). We use the classical Monte Carlo technique for performing a configurational averaging over $\eta$. It requires a few thousands of random realizations in order to obtain converged values. Our numerical searches show the presence of two kind of solutions, either $m_\perp^{\epsilon,2} \ne 0$, $m_\parallel^{\epsilon,2}=0$ (i.e., case I) or  $m_\perp^{\epsilon,2} = 0$, $m_\parallel^{\epsilon,2} \ne 0$ (i.e., case II), which is in accordance with previous discussions in the context of analytical perturbative analysis. 

Fig. \ref{fig_epsilon}(a) exhibits the results obtained by numerical analysis for the transverse magnetization, i.e. case I, with vanishing $y$-component and non-zero $x$-component, for $\epsilon/J=0.1$ and $\gamma=0.1$. When the temperature is high enough, the system does not magnetize, similarly to the case of an ordered system, i.e.  $\epsilon/J=0$.  However, below the critical temperature (i.e. if $\beta > \beta_{c,\perp}^{\epsilon,2}$), the system magnetizes in the direction transverse to the applied random field. We see that the critical temperature decreases in the presence of the disorder, i.e. that the critical point, $\beta_{c,\perp}^{\epsilon,2}$, shifts towards the right in presence of the random field. We also find excellent agreement between the exact numerical results and the approximate analytical expression of the transverse magnetization derived using a perturbative approach given in Eq.~(\ref{xy_disorder14}) (see inset of Fig.~\ref{fig_epsilon}(a)).

The numerical results for case II with vanishing $y$-component and non-zero $x$-component are shown in Fig.~\ref{fig_epsilon}(b) for $\epsilon/J=0.1$ and $\gamma=0.1$. The features of parallel magnetization are qualitatively similar to that of transverse magnetization. However, we find that the critical point, $\beta_{c,\parallel}^{\epsilon,2}$, as may be expected by now given the analytical results, shifts towards an even higher value compared to the case of transverse magnetization.  A closer examination of the Figs.~\ref{fig_epsilon}(a)-(b) show that the effect of disorder is more prominent in the parallel magnetization than in the tranverse one. This is confirmed by the expressions derived in the small-$m$ regime (see Eqs.~(\ref{xy_disorder14}) \& (\ref{xy_disorder16})).

The behavior of the transverse and the parallel magnetizations for a given $\epsilon$ and selectively chosen values of the anisotropy constant, $\gamma$, is 
demonstrated in Fig.~\ref{fig_epsilon_1}. We find that the inverse critical temperature, $\beta_{c,\perp}^{\epsilon}$, decreases with increasing $\gamma$ for the case when 
the system magnetizes in the direction which is transverse to the applied random field. The opposite happens when the system magnetizes in the direction which is parallel to 
the radom field. The insets of the Figs.~\ref{fig_epsilon_1}(a-b) show the critical temperatures $\beta_c$'s as functions of $\gamma$. The trends suggest that for highly anisotropic systems, 
the parallel magnetization would occur only at sufficiently low temperatures. High anisotropy favors transverse magnetization, i.e.  the system 
starts magnetizing in the transverse direction at comparetively higher temperatures.

\section{Order from disorder: Random field quantum XY MODEL in  the presence of AN ADDITIONAL UNIFORM FIELD}
\label{steady}
Until now, we have seen that the spontaneous magnetization in the system persists, albeit only in a restricted set of directions, even in the presence of a disordered field.  Is this still true when there is an additional constant field? In this section, we consider this question and show that not only does the spontaneous magnetization persist---disorder can now help one of the components of the magnetization to achieve an enhanced value compared to the ordered system. 

We first consider the case in which the ordered $XY$ model is subject to a constant magnetic field, $\vec{h}$. The mean-field Hamiltonian, $H_h$, governing the system in this case is given by
\begin{eqnarray}
\label{xy_disorder_con_field}
H_h=-\left(J\left((1+\gamma) m_x \sigma_x+(1-\gamma) m_y \sigma_y\right)+ \vec{h}\cdot \vec{\sigma}\right)
\end{eqnarray}
The constant field $\vec{h}$ lies in the $XY$ plane, i.e.~$\vec{h}=(h_x,h_y)=(h \cos{x}, h \sin{x})$ with magnitude $h$, where $0<h\leq1$, and phase $x$, with $-\pi/2\leq x \leq \pi/2$. In the presence of the constant field, the mean-field equation for the magnetization is obtained replacing $H$ by $H_h$ in Eq.~(\ref{xy_disorder}).  The system now has no critical temperature, as there is always a unique solution at any value of $\beta$. 
\\

Let us now investigate the effect of a random field, $\epsilon \vec{\eta}$, on the system. The mean-field Hamiltonian $H_{h,\epsilon}$ can be written as
\begin{eqnarray}
\label{xy_disorder_con_field}
H_{h,\epsilon}=-\left(J((1+\gamma) m_x \sigma_x+(1-\gamma) m_y \sigma_y)+ \vec{h} \cdot \vec{\sigma}+ \epsilon \eta \sigma_y\right),
\nonumber\\
\end{eqnarray}
where we assume the random field to be directed along the $y$-axis.  Replacing again $H$ by $H_{h,\epsilon}$ in Eq.~(\ref{xy_disorder}),  we obtain two coupled equations, which we solve to find $\vec{m}$. As may be expected, the solution for the magnetization is again unique.

\begin{figure}[t]
\includegraphics[angle=0,width=35mm]{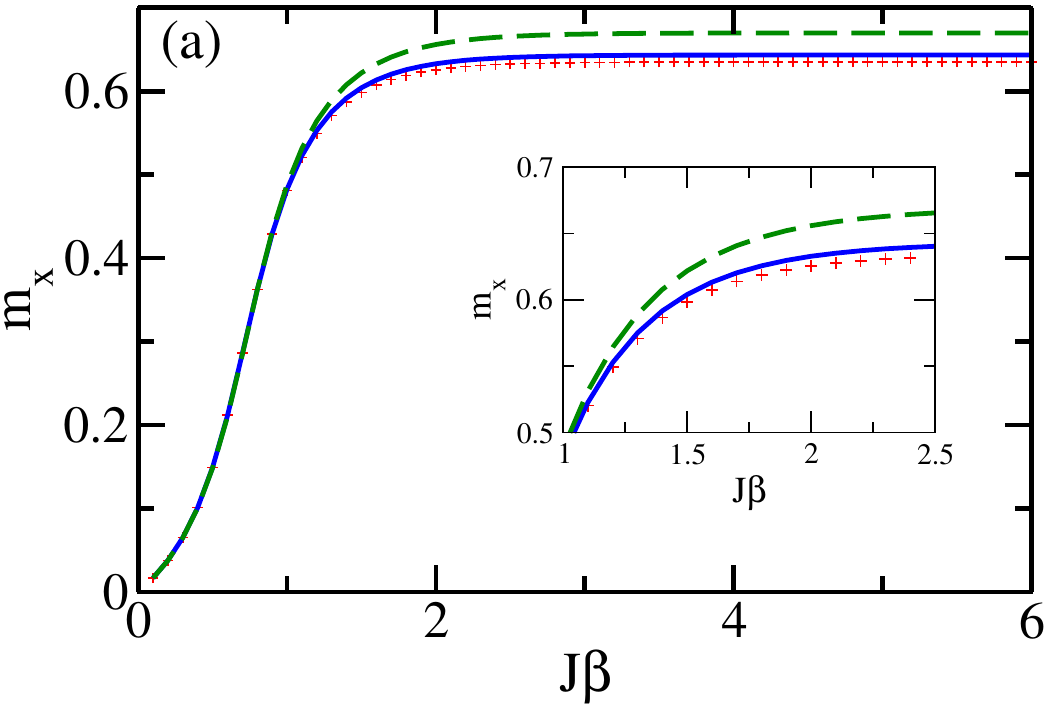}
\hspace{0.6cm}
\includegraphics[angle=0,width=35mm]{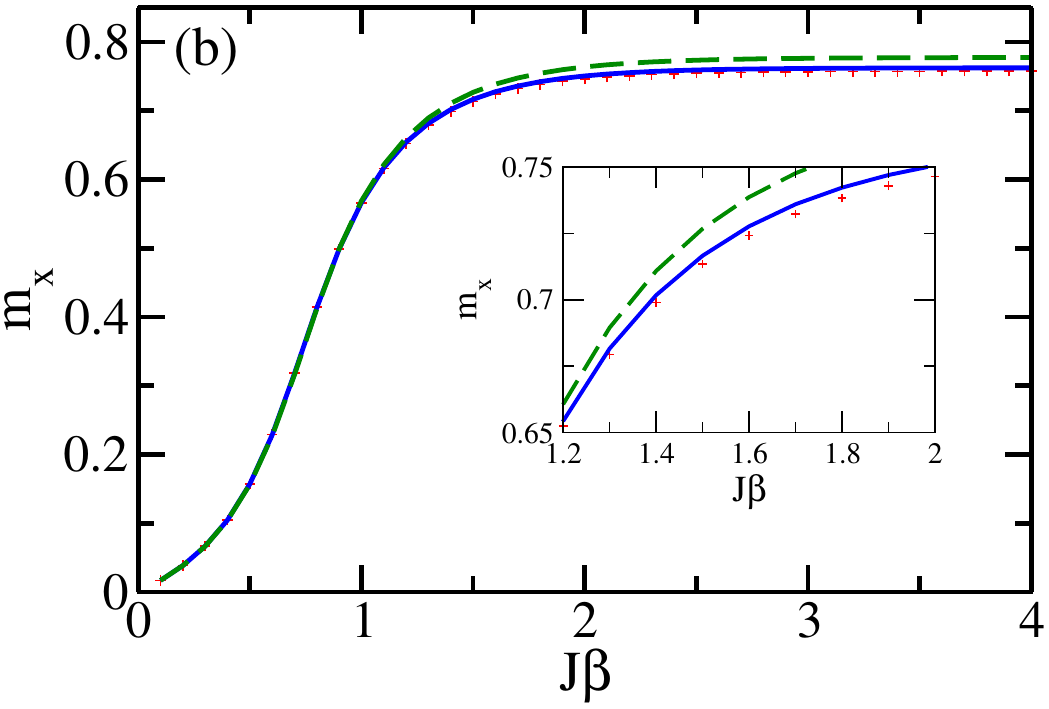}
\caption{ (Color online) Order from disorder. Effect of the random field in presence of a constant field. 
The $X$-component of magnetization,
$m_x$, as a function of $J \beta$ for (a) $\gamma=0.05$ and (b) $\gamma=0.1$. Red pluses show $m_x$ for the case
when the $XY$ model is subjected to a constant field $\vec{h}$ with $h/J =0.3$, and $x=\pi/3$. 
The blue solid and the green dashed lines represent $m_x$ when the system is treated with random field of strength $\epsilon/J=0.1$ and $\epsilon/J=0.2$, respectively, 
along with the constant field (the corresponding Hamiltonian is given in Eq.~(\ref{xy_disorder_con_field})). 
The insets show blow-ups of the same for a smaller range of $J\beta$. The enhancement of $m_x$ in presence of the disorder field uncovers a ``random field induced order". 
Comparing panels (a) and (b), we observe that for fixed \(h/J\) and \(x\), the order-from-disorder phenomenon appears as the anisotropy parameter is cranked down. 
All other quantities are dimensionless.}
\label{fig:steady-random}
\end{figure}

Our numerical calculations show that the magnetization $m$  and the $y$-component of the magnetization vector $m_y$ are reduced in length in the presence of disorder, i.e. when the system is governed by $H_{h,\epsilon}$ as compared to the ordered system described by $H_h$.   However, the $x$-component $m_x$ behaves in a very different manner. Depending upon the system parameters, $m_x$ can be both higher and lower than its value in the ordered system. In Fig.~\ref{fig:steady-random}, we exhibit the results in the particular example for the system with $h/J =0.3$, $x =\pi/3$, which demonstrate the random field induced enhancement of $m_x$ in the presence of  disorder for two different values of $\gamma$, signaling random-field-induced order, also known as ``order from disorder." Our numerical observations are further supported  by results obtained analytically via peturvative approach at low temperature. The details are discussed below.
 

\subsection{Perturbative analysis of the low-temperature magnetization} 
 The mean field equations in Eq.~(\ref{xy_disorder}) can be alternatively presented as
\begin{equation}
\label{rfcf1}
m_x=\frac{1}{\beta J (1+\gamma)} \frac{\partial\Gamma}{\partial m_x}
\end{equation}
and
\begin{equation}
\label{rfcf2}
m_y=\frac{1}{\beta J (1-\gamma)} \frac{\partial\Gamma}{\partial m_y},
\end{equation}
where 
\begin{equation}
\Gamma \equiv {\rm Av}_\eta\left[\log_e \Tr \exp(-\beta H_h)\right]
\end{equation}
 or in the disordered case, 
 \begin{equation}
 \Gamma \equiv {\rm Av}_\eta\left[\log_e \Tr \exp(-\beta H_{h,\epsilon})\right]
 \end{equation}
 depending on whether the governing  Hamiltonian is $H_h$ or $H_{h,\epsilon}$.
The symmetry of the Gaussian distribution of $\eta$ ensures that $\vec{m}$ is an even function of $\epsilon$. As a result, $\frac{d m_x}{d \epsilon}$ and $\frac{d m_y}{d \epsilon}$ vanish at $\epsilon = 0$. Starting with the Eqs.~(\ref{rfcf1}) and (\ref{rfcf2}), straightforward algebra leads to the following set of coupled equations:
\begin{eqnarray}
\label{rfcf3}
\nonumber 
\frac{d^2m_x}{d\epsilon^2}\left[1-\frac{1}{\beta J (1+\gamma)} \frac{\partial^2 \Gamma}{\partial m_x^2}\right]=
\nonumber\\
\frac{1}{\beta J (1+\gamma)} \left[ \frac{\partial^3\Gamma}{\partial^2\epsilon\partial m_x}
+\frac{\partial^2\Gamma}{\partial m_y \partial m_x} \frac{d^2m_y}{d \epsilon^2}\right],
\end{eqnarray}
and
\begin{eqnarray}
\label{rfcf4}
\frac{d^2m_y}{d\epsilon^2}\left[1-\frac{1}{\beta J (1-\gamma)} \frac{\partial^2 \Gamma}{\partial m_y^2}\right]=
\nonumber\\
\frac{1}{\beta J (1-\gamma)} \left[ \frac{\partial^3\Gamma}{\partial^2\epsilon\partial m_y}+\frac{\partial^2\Gamma}{\partial m_y \partial m_x} \frac{d^2m_x}{d \epsilon^2}\right],
\end{eqnarray}
where the total and partial derivatives are evaluated at $\epsilon = 0$. In order to evaluate $\frac{d^2m_x}{d\epsilon^2}$ and $\frac{d^2m_y}{d\epsilon^2}$, at $\epsilon=0$, we need to calculate the partial derivatives at $\epsilon=0$. For example, the expression for $\frac{1}{\beta J (1+\gamma)}\frac{\partial^2}{\partial \epsilon^2} \frac{\partial \Gamma}{\partial m_x}$ is
\begin{equation}
\label{rfcf5}
\frac{1}{\beta J (1+\gamma)} \frac{\partial \Gamma}{\partial m_x}=m_x=\frac{J m_x (1+\gamma)+h_x}{k_{\epsilon}{'}} \tanh(\beta k_{\epsilon}{'}).
\end{equation}
It follows that
\begin{eqnarray}
\label{partial-2}
\frac{1}{\beta J (1+\gamma)}\frac{\partial^2}{\partial \epsilon^2} \frac{\partial \Gamma}{\partial m_x}=
\nonumber\\
{\rm Av}_{\eta}\large[(J m_x (1+\gamma)+h_x) \big({\frac{-3 \beta \eta^2 (h_y+J m_y (1-\gamma)+\epsilon \eta)^2}{k_{\epsilon}{'}^4 \cosh^2(\beta k_{\epsilon}{'})}}
\nonumber\\
+\frac{\beta \eta^2}{k_{\epsilon}{'}^2 [\cosh^2(\beta k_{\epsilon}{'})]}-\frac{\eta^2 \tanh(\beta k_{\epsilon}{'})}{k_{\epsilon}{'}^3}
\nonumber\\
+\frac{3 \eta^2 (h_y+J m_y (1-\gamma)+\epsilon \eta)^2 \tanh(\beta k_{\epsilon}{'})}{k_{\epsilon}{'}^5}
\nonumber\\
-\frac{2 \beta^2 \eta^2 (h_y+J m_y (1-\gamma)+\epsilon \eta)^2 \tanh(\beta k_{\epsilon}{'})}{k_{\epsilon}{'}^3 \cosh^2(\beta k_{\epsilon}{'})} \big)\large].
\nonumber\\
\end{eqnarray}
Here $k_{\epsilon}{'}=\sqrt{((J (1+\gamma) m_x)+h_x)^2+((J (1-\gamma) m_y)+\epsilon \eta+h_y)^2}$.
Next, using the asymptotic expansion of the hyperbolic function $\tanh(\beta k_{\epsilon}{'}) \approx 1-2 \exp(-2 \beta k_{\epsilon}{'})$, we obtain  for the partial derivatives at $\epsilon=0$
\begin{equation}
\label{d2mx}
\frac{d^2m_x}{d \epsilon^2}|_{\epsilon=0}= \frac{1}{h^2} P \left(x,\frac{J}{h} \right) + O(e^{-\beta}),
\end{equation}
and
\begin{equation}
\label{d2my}
\frac{d^2m_y}{d \epsilon^2}|_{\epsilon=0}=\frac{1}{h^2} Q \left(x,\frac{J}{h} \right) + O(e^{-\beta}),
\end{equation}
\begin{figure}[t]
\vspace*{+.2cm}
\includegraphics[angle=270,width=42mm]{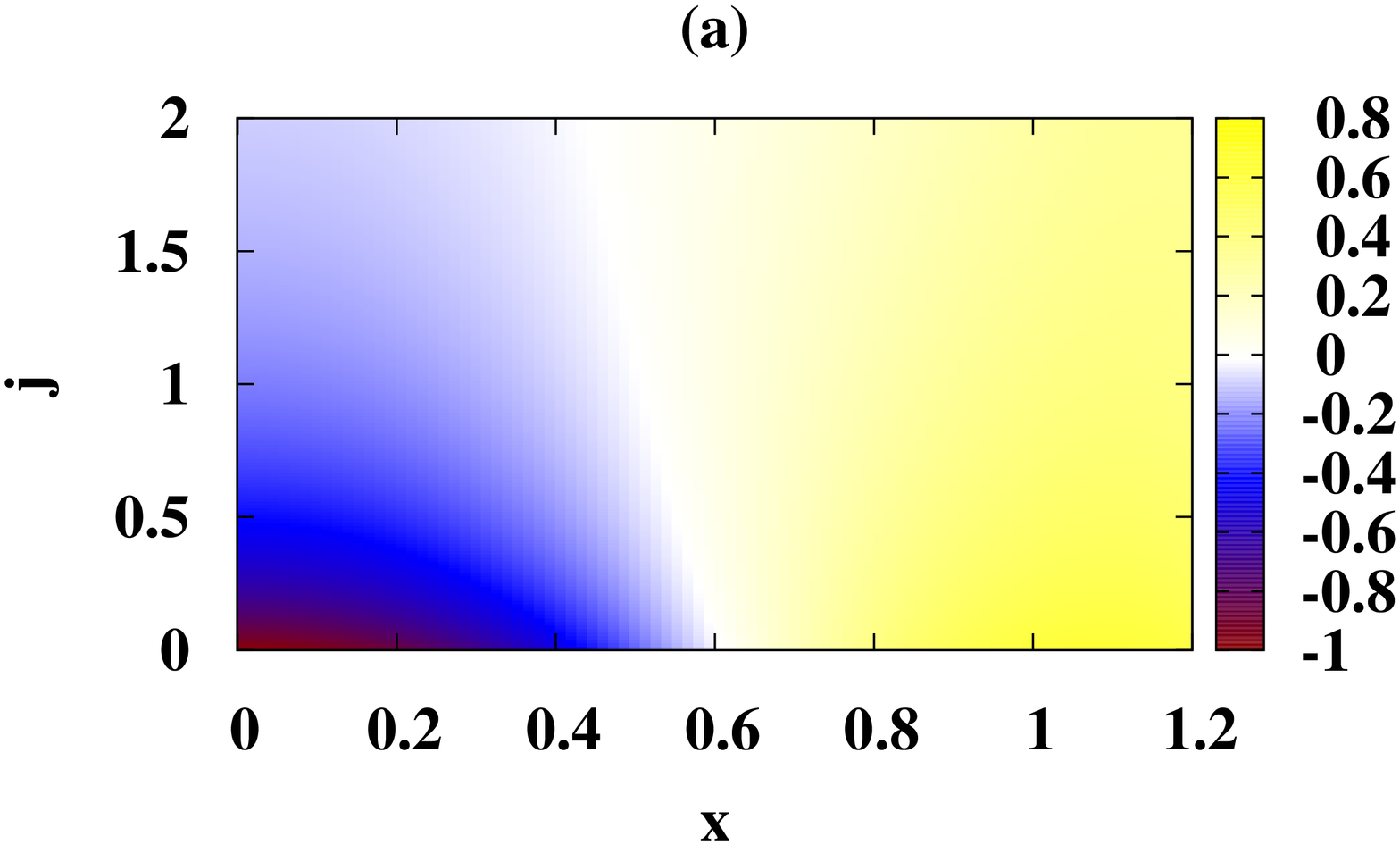}
\includegraphics[angle=270,width=42mm]{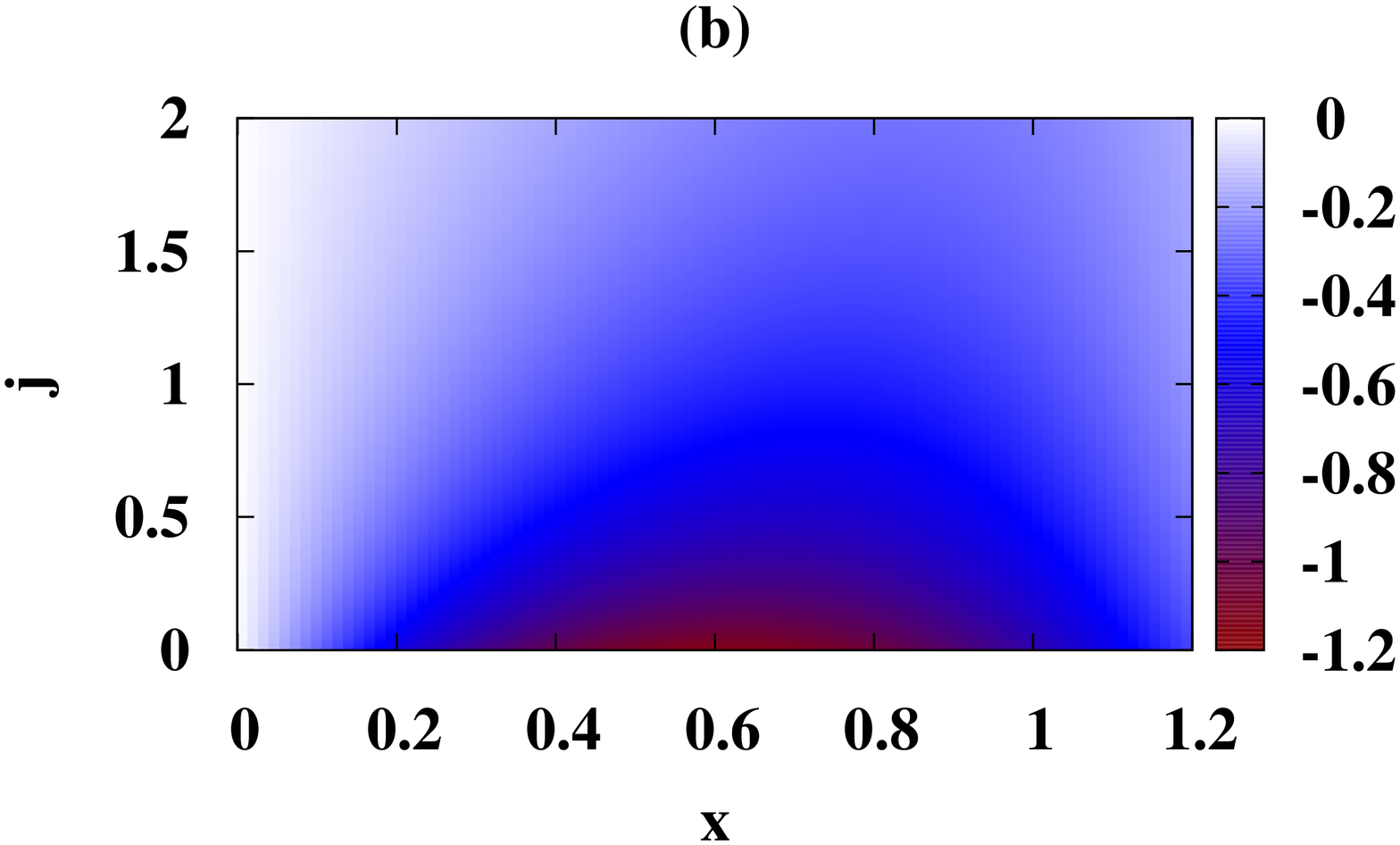}\\
\includegraphics[angle=270,width=42mm]{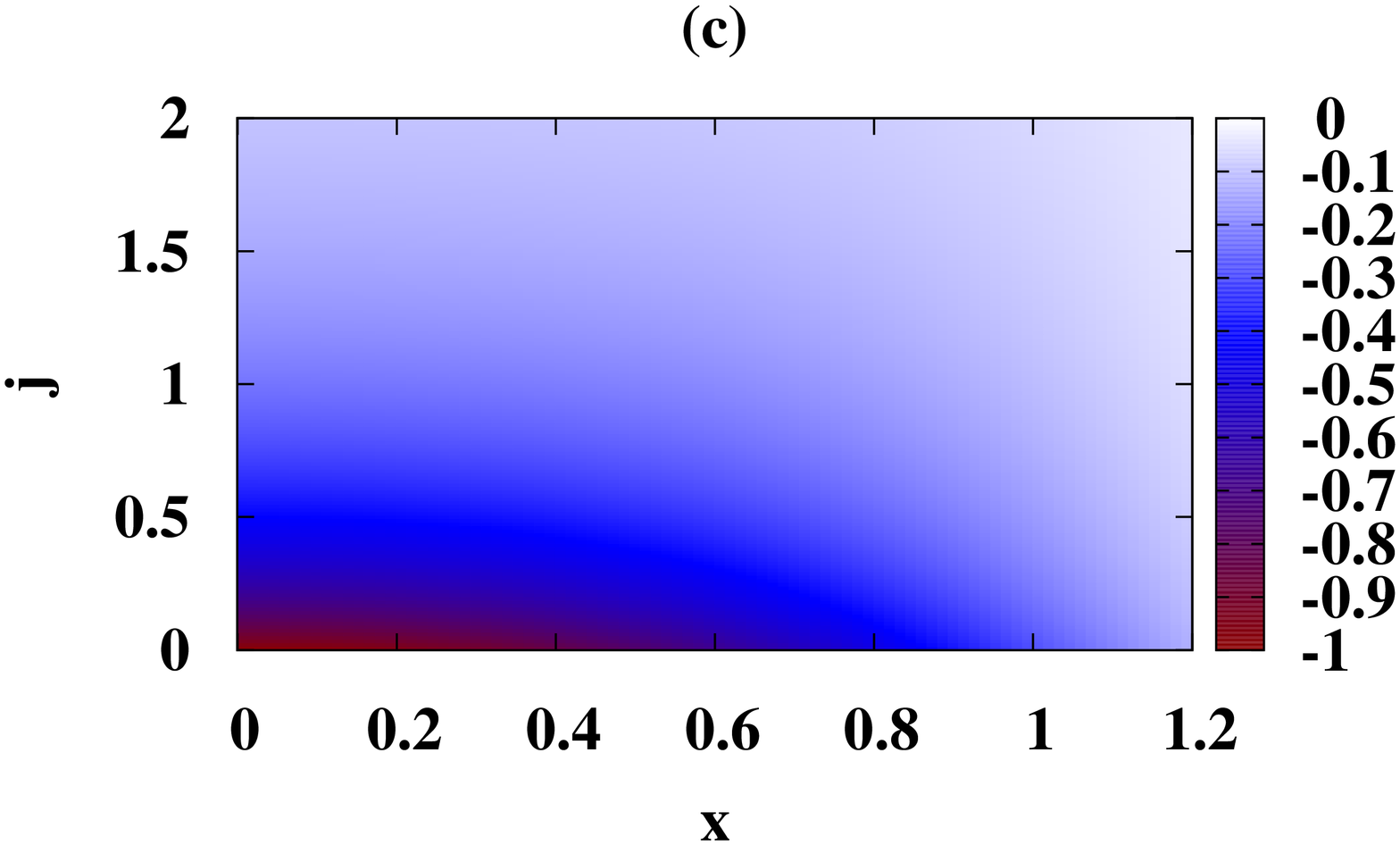}
\includegraphics[angle=270,width=42mm]{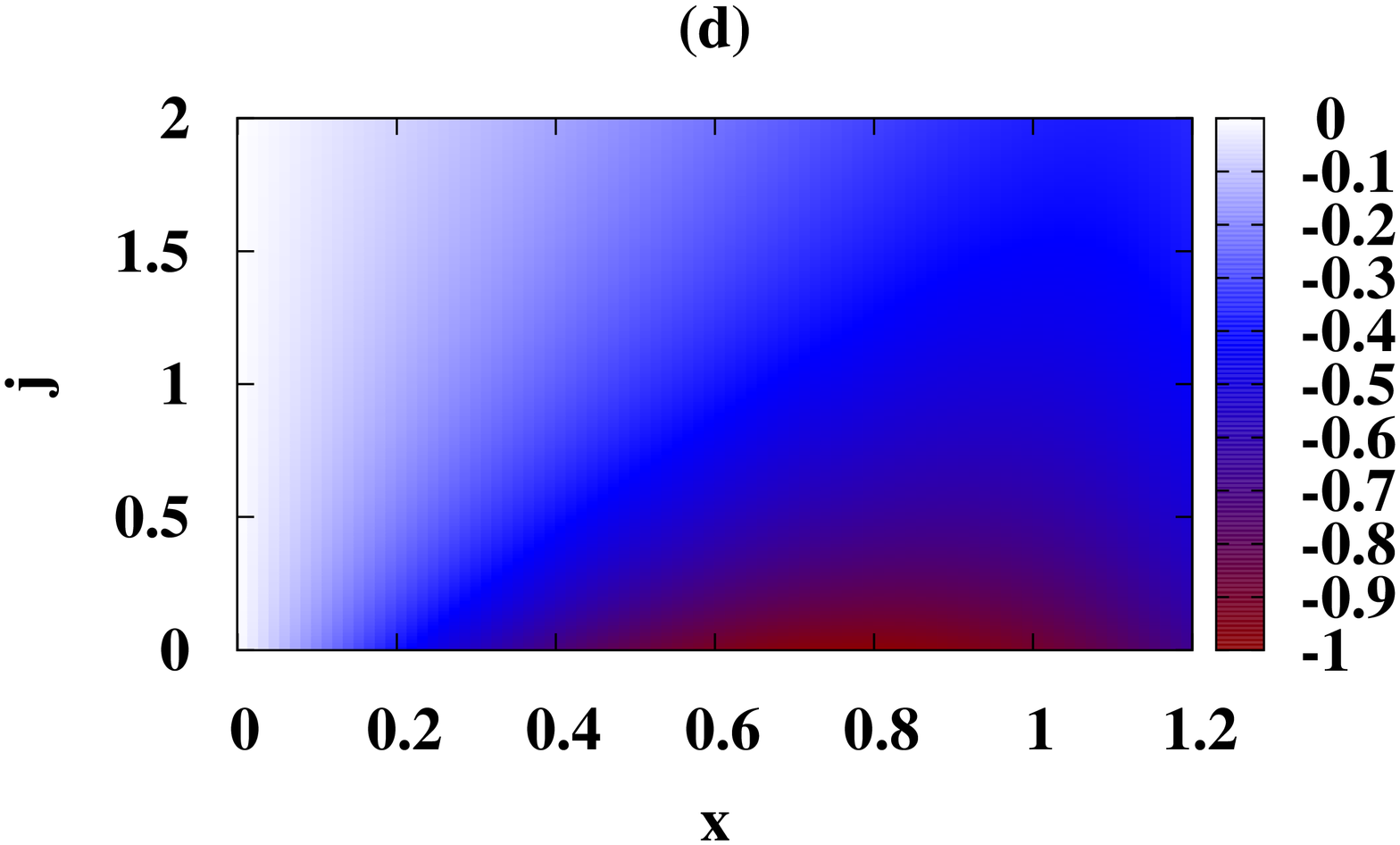}
\vspace*{0.2cm}
\caption{(Color online) The span of order-from-disorder in parameter space. Plots of the functions (a) $P(x,j)$, (b) $Q(x,j)$, (c) $R(x,j)$, and (d) $S(x,j)$  with respect to $x$ and $j=J/h$. Note that there are ranges of the $(x,j)$, for which the function $P(x,j)$ is positive signaling order from disorder in the system described by the Hamiltonian $H_{h,\epsilon}$. However, this is not true for $Q(x,j)$ and $R(x,j)$, which are negative for the entire range of $x$ and $j$. The fact that $S(x,j)$ is also negative in the entire range implies that in the presence of disorder, the magnetization vector moves away from the direction of the applied random field. All other quantities are dimensionless. Here $\gamma=0.1$.}
\label{PandQ}
\end{figure}
where the functions $P$ and $Q$ are given by (for $j=J/h$)
\begin{equation}
\label{P-new}
P \left(x,j \right) = \frac{A E+B C}{D E-C C{'}}
\end{equation}
and
\begin{equation}
\label{Q-new}
Q \left(x,j \right) = \frac{A C{'}+B D}{D E-C C{'}}.
\end{equation}
Here
\begin{eqnarray}
\label{A-new}
A(x,j)=a \cos x (\frac{3 b^2 \sin^2 x}{k{'^5}}-\frac{1}{k{'^3}}),
\end{eqnarray}
\begin{eqnarray}
\label{B-new}
B(x,j)=b \sin x \left(\frac{3 b^2 \sin^2 x}{k{'^5}}-\frac{3}{k{'^3}}\right),
\end{eqnarray}
\begin{eqnarray}
\label{C-new}
C(x,j) = -\frac{J a b \cos x \sin x}{k{'^3}} (1-\gamma),
\end{eqnarray}
\begin{eqnarray}
\label{C-dash-new}
C{'}(x,j) = -\frac{J a b \cos x \sin x}{k{'^3}} (1+\gamma),
\end{eqnarray}
\begin{eqnarray}
\label{D-new}
D(x,j) = 1-\frac{J (1+\gamma)}{k{'}} (1-\frac{a^2 \cos x}{k{'^2}}),
\end{eqnarray}
and
\begin{eqnarray}
\label{E-new}
E(x,j) = 1-\frac{J (1-\gamma)}{k{'}} (1-\frac{b^2 \sin x}{k{'^2}}),
\end{eqnarray}
with $a=1+j (1+\gamma)$, $b=1+j (1-\gamma)$, and $k{'}=\sqrt{a^2 \cos^2 x+b^2 \sin^2 x }$. The positivity of $P(x,j)$ implies that order from disorder phenomenon occurs.

As is clear from Fig.~\ref{PandQ}(a), there exists a region in the parameter space $(x,j)$, for which $P(x,j)>0$ which confirms that the quenched averaged $X$-component $m_x$ of the magnetization is enhanced by the presence of disorder.  This does not hold for the quenched averaged $Y$-component, $m_y$, which is reduced in length in the presence of disorder (see Fig.~\ref{PandQ}(b)).

To further investigate the effect of disorder on the length $m$ and phase $\phi_1$ of the magnetization, we expand $\tan\phi_1 = \frac{m_y}{m_x}$ as
\begin{eqnarray}
\label{phase-1}
\tan\phi_1={\frac{m_y}{m_x}}\big|_{\epsilon=0}+\epsilon^2 \frac{d^2}{d\epsilon^2}\left(\frac{m_y}{m_x}\right)\big|_{\epsilon=0}+O(\epsilon^4),
\end{eqnarray}
with
\begin{eqnarray}
\label{phase-2}
 \frac{d^2}{d\epsilon^2}\left(\frac{m_y}{m_x}\right)\big|_{\epsilon=0}=\frac{m_x \frac{d^2m_y}{d\epsilon^2}-m_y \frac{d^2m_x}{d\epsilon^2}}{m_x^2}\big|_{\epsilon=0}
\nonumber\\
=\frac{1}{m_x^2\big|_{\epsilon=0}}\frac{1}{h^2} S(x,j)+O(e^{-\beta}),
\end{eqnarray}
where
\begin{equation}
\label{S-new}
S(x,j)=Q(x,j) \cos x-P(x,j) \sin x.
\end{equation}
$S(x, j)$ is negative for all $x$ and $J/h$ (see Fig.~\ref{PandQ}(d)) implying that the phase always shift towards the $X$-axis in the presence of the random field.

The square of the length of the magnetization, when similarly expanded, is given by
\begin{equation}
\label{magnitude}
m_x^2+m_y^2=(m_x^2+m_y^2)\big|_{\epsilon=0}+2\epsilon^2\left(R(x,j)+O(e^{-\beta})\right),
\end{equation}
where
\begin{equation}
\label{R-new}
R(x,j)=(P \cos x+Q \sin x)\big|_{\epsilon=0}.
\end{equation}
As seen in Fig.~\ref{PandQ}(c), $R(x,j)$ is negative regardless of the choice of parameters, i.e., the length of the magnetization decreases in the presence of the disorder. Note that the analytical results are in agreement with the numerical evidence presented above. It is worth mentioning here that the analytical results are valid for small $\epsilon$ and large $\beta$. The difference between the magnetization in the disordered system and the ordered system, as obtained analytically, is of the order of $\epsilon^2$. Comparison of these analytical results with the numerical ones is valid only when the same difference, obtained numerically, has precision of order $\epsilon^2$.

\vspace{2mm}
\section{Generalization to arbitrary spins and scaling of critical temperature}
\label{general}
In this section, our aim is to investigate $d$-dimensional lattices where the occupant of each lattice site is a quantum spin with arbitrary spin angular momentum. Here we restrict ourselves to the $XX$ model. For our purposes, it is necessary to treat the half-integer and integer spins separately. In the following subsections, we derive the generalized expressions for the scaling of the magnetization and critical temperature for both cases.

\subsection{Half-integer spins}
The mean-field equations for a general half-integer spin ${n+1 \over 2}$, ($n = 1,3, \dots$) are
\begin{equation}
\label{gen11}
m_x={\rm Av}_\eta  \left[\frac{J m_x}{k} \frac{\sum_{p=0}^n (2 p+1) \sinh(2 p+1) \beta k}{\sum_{p=0}^n \cosh(2 p+1) \beta k}\right],
\end{equation}
\begin{equation}
\label{gen12}
m_y={\rm Av}_\eta \left[\frac{J m_y+\epsilon \eta}{k} \frac{\sum_{p=0}^n (2 p+1) \sinh(2 p+1) \beta k}{\sum_{p=0}^n \cosh(2 p+1) \beta k}\right],
\end{equation} 
where $k=\sqrt{J^2 m_x^2+(J m_y+\epsilon \eta)^2}$.\\

Finding the magnetization $\vec{m}$ requires simultaneous  solution of the coupled set of Eqs.~(\ref{gen11}) and (\ref{gen12}), i.e., finding the common zeros of the following two functions:
\begin{eqnarray}
\label{gen13}
F_{x}^{\epsilon,n}(\vec{m})=
\nonumber\\
{\rm Av}_\eta \left[\frac{J m_x}{k} \frac{\sum_{p=0}^n (2 p+1) \sinh(2 p+1) \beta k}{\sum_{p=0}^n \cosh(2 p+1) \beta k}\right]-m_x, 
\end{eqnarray}
\begin{eqnarray}
\label{gen14}
F_{y}^{\epsilon,n}(\vec{m})=
\nonumber\\
{\rm Av}_\eta \left[\frac{J m_y+\epsilon \eta}{k} \frac{\sum_{p=0}^n (2 p+1) \sinh(2 p+1) \beta k}{\sum_{p=0}^n \cosh(2 p+1) \beta k}\right]-m_y.\nonumber\\
\end{eqnarray}

\begin{figure}[t]
\vspace*{+.2cm}
\includegraphics[angle=0,width=70mm]{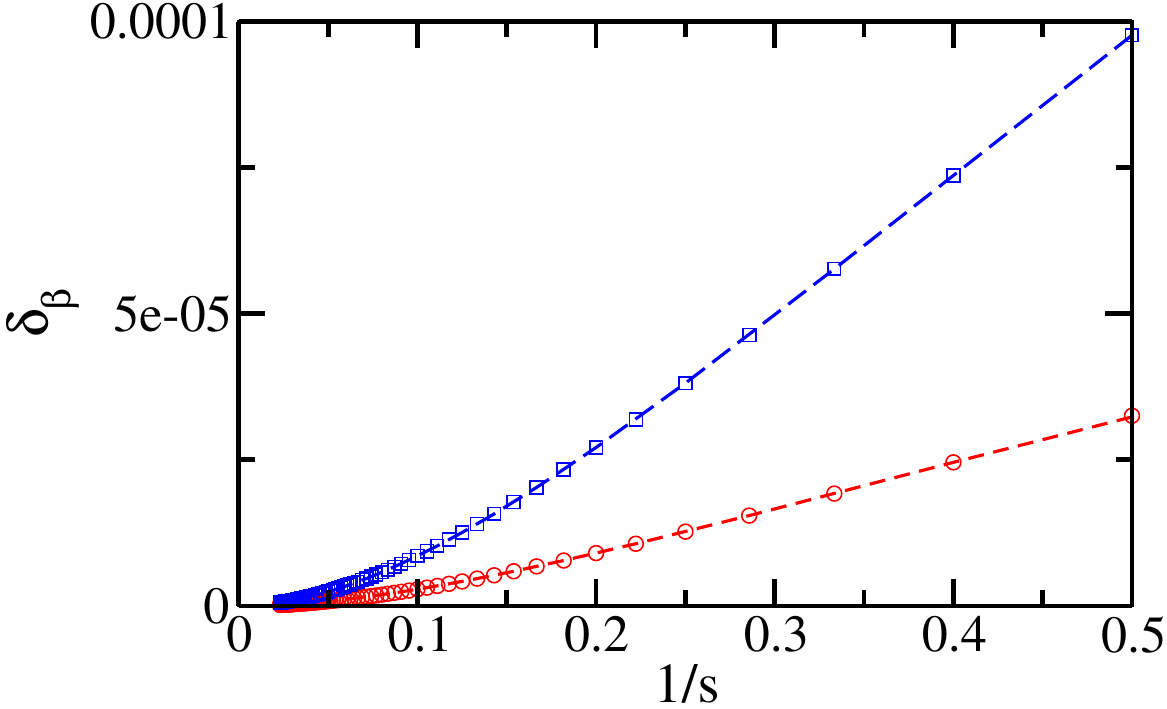}
\vspace*{0.2cm}
\caption{ (Color online.) $\delta_\beta$ as function of $1/s$ for the transverse (red circles) and parallel (blue squares) magnetizations for $\epsilon/J=0.05$. The lines serve as guides to the eye. All quantities are dimensionless.}
\label{fig:effbeta}
\end{figure}
The Taylor expansion in $\epsilon$, followed by the expansion in $m$, of the functions given in Eqs.~(\ref{gen13}) and (\ref{gen14}), around $\epsilon=0$ and $m=0$, gives
\begin{eqnarray}
\label{gen15}
F_{x}^{\epsilon,s}(\vec{m})=\frac{1}{45} [-45+60 J \beta s (s+1)-
\nonumber\\
8 J s (s+1) (2 s^2+2 s+1) \beta^3 \epsilon^2] m \cos\phi_1+
\nonumber\\
\frac{1}{3!} \frac{16}{315} \beta^3 J^3 [s (s+1) (-21 (2 s^2+2 s+1)+
\nonumber\\
2 \beta^2 \epsilon^2 (4 s^2+2 s+1) (4 s^2+6 s+3))] m^3 \cos\phi_1+O(m^5), 
\nonumber\\
\end{eqnarray} 
and
\begin{eqnarray}
\label{gen16}
F_{y}^{\epsilon,s}(\vec{m})=\frac{1}{15} [-15+20 J \beta s (s+1)-
\nonumber\\
8 J s (s+1) (2 s^2+2 s+1) \beta^3 \epsilon^2] m \sin\phi_1+
\nonumber\\
\frac{1}{3!} \frac{16}{315} \beta^3 J^3 [s (s+1) (-21 (2 s^2+2 s+1)+
\nonumber\\
10 \beta^2 \epsilon^2 (4 s^2+2 s+1) (4 s^2+6 s+3))] m^3 \sin\phi_1+O(m^5),\nonumber\\
\end{eqnarray}
where $s= n+1/2$ with $n=0,1,2$,\ldots .
$\phi_1$ has two allowed  values: $\pi/2$ (system magnetizes in the direction parallel to the disordered field) and $0$ (system magnetizes in the direction transverse to the disordered field). For transverse magnetization, $F_y^{\epsilon,n}(\vec{m})$ vanishes and two non-trivial solutions solely come from Eq.~(\ref{gen15}) as
\begin{widetext}
\begin{eqnarray}
\label{gen17}
\label{beta-perp-m}
m_\perp^{\epsilon,s}=\pm \sqrt{\frac{21}{8}} \sqrt{\frac{[45-60 J \beta s (s+1)+8 J s (s+1) (2 s^2+2 s+1) \beta^3 \epsilon^2]}{J^3 \beta^3 s (s+1) [-21 (2 s^2+2 s+1)+2 \beta^2 \epsilon^2 (4 s^2+2 s+1) (4 s^2+6 s+3)]}}.
\end{eqnarray}
\end{widetext}
The critical point can now be easily obtained by setting $m_\perp^{\epsilon,s}=0$ in Eq.~(\ref{gen17}). We get
\begin{eqnarray}
\label{gen18}
8 s (s+1) (2 s^2+2 s+1) J \beta^3 \epsilon^2-60 s (s+1) J \beta+45=0,
\nonumber\\
\end{eqnarray}
which gives
\begin{equation}
\label{beta-perp-halfint}
\beta_{c,\perp}^{\epsilon,s}=\frac{3}{4 J s (s+1)}+\frac{9}{160} \frac{(2 s^2+2 s+1)}{J^3 s^3 (s+1)^3} \epsilon^2,
\end{equation}
The critical temperature decreases with the increase of number of spin. The shift in critical temperature is of the order of $\epsilon^2$ for all spins. Note that the generalized expressions for the scaling and for the critical temperature for the pure system with a transverse magnetization can be obtained simply by putting  $\epsilon=0$ in Eqs.~(\ref{beta-perp-m}) and (\ref{beta-perp-halfint}), respectively. 

In order to find the expressions for the parallel magnetization, we put $\phi_1=\pi/2$ in the Eqs.~(\ref{gen15}) and (\ref{gen16}). In this case also, the right-hand side of Eq.~(\ref{gen15}) vanishes to leading order, while Eq.~(\ref{gen16}) has two nontrivial solutions, given by:
\begin{widetext}
\begin{eqnarray}
\label{gen19}
m_\parallel^{\epsilon,s}=\pm \sqrt{\frac{63}{8}} \sqrt{\frac{15-20 J s (s+1) \beta+8 J s (s+1) (2 s^2+2 s+1) \beta^3 \epsilon^2}{J ^3 \beta^3 s (s+1) [-21 (2 s^2+2 s+1)+10 \beta^2 \epsilon^2 (4 s^2+2 s+1) (4 s^2+6 s+3)]}}.
\end{eqnarray}
\end{widetext}
The critical point can be obtained by considering $m_\parallel^{\epsilon,s}=0$ in Eq.~(\ref{gen19}) and we obtain
\begin{equation}
\label{gen20}
\beta_{c,\parallel}^{\epsilon,s}=\frac{3}{4 J s (s+1)}+\frac{27}{160} \frac{(2 s^2+2 s+1)}{J^3 s^3 (s+1)^3} \epsilon^2
\end{equation}
The generalized expressions of the scaling and the critical temperature for the pure system with a parallel magnetization can again be obtained by putting  $\epsilon=0$ in Eqs.~(\ref{gen19}) and (\ref{gen20}), respectively. However, the shift in the critical temperature due to the random field is in this case bigger than in the transverse case and hence the effect of the  disorder is more prominent in the parallel case similarly to what was seen in section (\ref{near}) without constant field.

\subsection{Integer spins}
The generalized mean-field equations for the system with integer spin ${n \over 2}$, $n$ even, are given by
\begin{equation}
\label{gen1}
m_x={\rm Av}_\eta \left[\frac{J m_x}{k} \frac{\sum_{p=1}^n2 p(e^{2 p \beta k}-e^{-2 p \beta k})}{1+\sum_{p=1}^n(e^{2 p \beta k}+e^{-2 p \beta k})}\right],
\end{equation}
\begin{equation}
\label{gen2}
m_y={\rm Av}_\eta \left[\frac{J m_y+\epsilon \eta}{k} \frac{\sum_{p=1}^n2 p(e^{2 p \beta k}-e^{-2 p \beta k})}{1+\sum_{p=1}^n(e^{2 p \beta k}+e^{-2 p \beta k})}\right],
\end{equation} 
where $k=\sqrt{J^2 m_x^2+(J m_y+\epsilon \eta)^2}$. Now in order to find the magnetization $\vec{m}$ we have to solve the coupled set of 
Eqs.~(\ref{gen1}) and (\ref{gen2}). As one can expect from the previous discussions, there are two different kinds of magnetizations---the transverse magnetization, $m_\perp^{\epsilon,s}$, and the parallel magnetization, $m_\parallel^{\epsilon,s}$.  To derive the critical scaling for this case, we follow a Taylor expansion  method, similar to the one used for the half-integer spin case.  The final expressions for $m_\perp^{\epsilon,s}$,  $m_\parallel^{\epsilon,s}$, and the associated critical temperatures are given by the set of Eqs.~(\ref{beta-perp-m}),(\ref{beta-perp-halfint}),(\ref{gen19}),(\ref{gen20}) with $s=n$, where $n=1,2,3,\ldots$ 

Therefore, we again obtain corrections of order $\epsilon^2$ to the critical temperature for all the integer spin systems. Again, the effect of disorder is more pronounced in the parallel magnetization case than in the transverse case.

\subsection{Critical temperature versus spin quantum number}
 In order to study the effect of disorder as a function of $s$, we define the dimensionless quantity $\delta_{\beta}$, given by
\begin{equation}
\label{delta-beta}
\delta_{\beta}=\frac{\beta_{c}^{\epsilon,s}-\beta_{c}^{0,s}}{\beta_{c}^{0,s}}. 
\end{equation}
$\delta_{\beta}$ is shown as a function of $1/s$ in Fig.\ref{fig:effbeta} for $\epsilon/J=0.05$.  We find that the shift in the critical temperature caused by the random field decreases with increasing spin quantum number.

\section{Conclusions}
\label{summary}
We considered the quantum spin-$1/2$ $XY$ model within the mean-field approximation and showed that the spontaneous magnetization persists in the system with the introduction of a unidirectional quenched disordered field, albeit it is smaller than in the pure system. Below a certain critical temperature, the magnetization occurs in specific directions, either parallel or transverse to the disordered field. The critical temperatures and the magnitude of the magnetization decrease with increasing strength of the disorder. We found perturbative expressions for scaling of the magnetization and the expressions for the scaling of the critical temperatures at which the system magnetizes. We also performed numerical simulations to obtain the behavior of magnetization for various values of the temperature, disorder strength and anisotropy parameter,  which match with the perturbative calculations for small disorder values.  Moreover, we extended our analysis to arbitary values of  (half-integer or integer) spin. We found that the decrease in the length of the magnetization due to the random field is of the order of the square of the strength of the disorder for all values of spin. The magnitude of magnetizations in the disordered systems decrease faster with increasing spin quatum number and the system requires lower temperature to magnetize when the spin quantum number increases, implying that the effect of disorder increases with increasing spin quantum number. In addition, we studied the random field quantum spin-$1/2$ $XY$ model with an additional constant field, for which we showed a random-field-induced ordering in the component of magnetization transverse to the disordered field. 

\acknowledgments
A.B. acknowledges the support of the Department of Science and Technology (DST), Govt.~of India, through the award of an INSPIRE fellowship. M.L. acknowledges Spanish MINECO Project  FOQUS (FIS2013-46768), ERC AdG OSYRIS, EU IP SIQS, EU STREP EQuaM, and EU FETPROACT QUIC.  J.W. was partially supported by NSF grant DMS 131271.

\end{document}